\documentclass[%
 reprint,
 amsmath,amssymb,
 prb,
 floatfix,
 groupedaddress,
]{revtex4-2}

\usepackage{graphicx}
\usepackage{dcolumn}
\usepackage{multirow}
\usepackage{bm}
\usepackage{bbold}

\begin{document}


\title{Optical Activity of Solids from First Principles}

\author{Xiaoming Wang}
\email{xiaoming.wang@utoledo.edu}
\affiliation{%
 Department of Physics and Astronomy, and Wright Center for Photovoltaics Innovation and Commercialization, The University of Toledo, Toledo, OH 43606, US
}%

\author{Yanfa Yan}%
\email{yanfa.yan@utoledo.edu}
\affiliation{%
 Department of Physics and Astronomy, and Wright Center for Photovoltaics Innovation and Commercialization, The University of Toledo, Toledo, OH 43606, US
}%

\date{\today}

\begin{abstract}
Within the framework of independent particle approximation, the optical activity tensor of solids is formulated as from different contributions: the magnetic dipole, electric quadrupole, and band dispersion terms. 
The first two terms have similar counterparts in the theory of finite systems, while the last term is unique for crystals. 
The magnetic dipole and electric quadrupole transition moments are calculated with a sum-over-states formulation. 
We apply the formulation to calculate and analyze the optical rotation of elemental tellurium and the circular dichroism of $(6,4)$ carbon nanotube. 
Decomposed optical activity into different contributions are discussed. 
The calculated spectra agree well with experiments.
As a showcase of achiral crystals, we calculate the optical activity of wurtzite GaN.
\end{abstract}

\maketitle

\section{Introduction}

Optical activity manifests itself as optical rotation (OR), the rotation $\rho$ of the plane of polarization of linearly polarized light, and circular dichroism (CD), the differential absorption of left- and right-circularly polarized light that leads to a change of the ellipticity $\theta$, as a result of light interacting with optically active media. 
Since most of the optically active media are chiral, chirality and optical activity are closely related. 
Optical activity is widely studied for chiral molecules which form the basis of biological, pharmaceutical, and life science~\cite{Barron2004}.
For crystalline solids, however, research is almost limited to OR~\cite{Jerphagnon1976, Nomura1960, Stolze1977, Ades1975, Cooper1969, Skalwold2015, Fukuda1975}, whereas CD is seldom due to the small effect~\cite{Stolze1977,Saeva1977}.
Only until recently that the excitonic effects are greatly enhanced in low-dimensional materials, can the CD be unambiguously observed. 
Typical examples are chiral carbon nanotubes (CNTs)~\cite{Wei2016,Peng2007,Green2009} and chiral hybrid halide perovskites~\cite{Long2018,Long2020} which hold potential for future spintronics.

The theory of optical activity of molecules has long been well established~\cite{Barron2004,Condon1937,Kirkwood1937,Rosenfeld1929}. 
It is based on the multipole theory of charge and current distributions. 
Both the magnetic dipole and electric quadrupole can contribute to the optical activity. 
However, the electric quadrupole term are usually considered to be small and thus neglected~\cite{Barron2004}. 
Hence, the magnetic dipole transition moment plays an important role in calculating and mechanistically analyzing the optical activity of molecules. 
\textit{Ab initio} calculations of the magnetic dipole transition moments, thus OR as well as CD, are now routinely available for the  quantum chemistry community because the quantum magnetic dipole operator \textbf{m} is well behaved for finite systems:
\begin{equation}
    \label{eq:one}
    \mathbf{m} = \frac{e}{2m} \mathbf{r} \times \mathbf{p}
\end{equation}
where \textit{e} and \textit{m} are the charge and mass of electron, \textbf{r} and \textbf{p} are the position and momentum operators, respectively. 
For crystals with periodic boundary condition, \textbf{m} in the above form is ill-defined due to the unbound operator \textbf{r}. 
Inserting a complete set between \textbf{r} and \textbf{p} in Eq.~(\ref{eq:one}) was used to calculate the magnetic dipole transition moments and thus the CD of chiral CNTs~\cite{Hidalgo2009,Sanchez2010}. 
In their formulation, the magnetic dipole transition moment is not Hermitian with respect to the band index, i.e., $\mathbf{m}_{ab} \neq \mathbf{m}_{ba}^\dagger $. 
Moreover, the electric quadrupole term and another band dispersion term which we will see later are neglected. 
Very recently, linear-response method was employed to directly evaluate the magnetic dipole and electric quadrupole transition moments with Gaussian basis for periodic systems~\cite{Rerat2021,Balduf2022}. 
Although the magnetic dipole and electric quadrupole matrix are enforced to be Hermitian through symmetrization, the obtained magnetic dipole and electric quadrupole contributions to the optical activity are individually origin-dependent.  
Also, the band dispersion contributions are not included. 

Alternatively, optical activity is a manifestation of the spatial dispersion, appearing at the first order in the expansion of the dielectric function in powers of light wavevector \textbf{q}. 
In this regard, formulation of the optical activity based on the expansion of the interaction Hamiltonian to the first order of \textbf{q} has been derived and applied to evaluate the OR of several simple chiral crystals~\cite{Zhong1992,Zhong1993,Jonsson1996,Malashevich2010prb,Tsirkin2018}.
This formulation is accurate in the sense that it includes all the necessary terms to the first order of \textbf{q}. 
However, applications for calculating the CD spectra have never been reported.
With this formulation, different contributions are mixed which may restrict some mechanistic analysis and understanding of the physics. 
In addition, available implementations are based on maximally-localized Wannier functions~\cite{Tsirkin2018}. 
As the conduction states are concerned for evaluating the transition moments, the wannierisation process itself is challenging for complex systems and broader energy range. 
In this manuscript, based on the work of Malashevich and Souza~\cite{Malashevich2010prb}, we formulate the optical activity by different contributions, i.e., the magnetic dipole term, the electric quadrupole term, and the band dispersion term. 
The former two contributions have direct connection with that of molecular systems, while the latter is unique for crystals. 
We implement our formulation as a postprocessing procedure based on the outputs of \textsc{vasp} code~\cite{Kresse1996cms,Kresse1996prb}.
For applications, we calculate and analyze the OR of Te and CD of chiral $(6,4)$ CNT. 
Good agreements with experiments are obtained.
Finally, we also discuss the optical activity of achiral crystals.

\section{Theory}

The spectra of OR and CD, related to each other with Kramers-Kr\"{o}nig transformation, can be calculated as~\cite{Yu2020,Tsirkin2018,Landau1984}
\begin{equation}
    \label{eq:two}
    \rho(\omega)+ i \theta(\omega) = \frac{\omega^2}{2c^2}\gamma(\omega)
\end{equation}
where $i$ is the imaginary unit, $\omega$ and \textit{c} are the frequency and speed of light, respectively. 
Here $\rho$ and $\theta$ share the same unit of degree per length.
In CD experiments, the ellipticity angle can be obtained by multiplying $\theta$ by the sample length.
$\gamma(\omega)=-i\partial \varepsilon^A (\omega, \mathbf{q})/\partial \mathbf{q}$ is the optical activity tensor where $\varepsilon^A (\omega, \mathbf{q})$ is the asymmetric part of the dielectric function. 
The \textbf{q} dependence indicates the spatial dispersion. 
In the independent particle framework, the dielectric function tensor can be calculated as~\cite{Adler1962} 
\begin{eqnarray}
    \label{eq:three}
    \varepsilon_{ij}(\omega,\mathbf{q})=&&\left(1-\frac{\omega_p}{\omega^2}\right)\delta_{ij} \nonumber \\
    &&+\frac{e^2}{\Omega \varepsilon_0 \omega^2 \hbar} \sum_{\sigma} \frac{\mathcal{K}_{\sigma,ij}(\mathbf{q}) f_\sigma (\mathbf{q})}{\omega-\omega_{\sigma}(\mathbf{q})+i \eta} 
\end{eqnarray}
where $\omega_p$ is the plasma frequency, $\Omega$ is the unit cell volume, $\varepsilon_0$ is the vacuum permittivity, $\hbar$ is the reduced Planck constant, $\eta$ is the broadening width.
The summation index $\sigma \equiv \{\mathbf{k}nm\}$ where $\mathbf{k}$ is the crystal momentum, \textit{n} and \textit{m} are band indexes.
In the manuscript, the subscripts $i,j,l,u,v \in \{1,2,3\}$ denote the Cartesian indices.
The occupation factor difference $f_\sigma(\mathbf{q})=f_{n\mathbf{k}+\frac{\mathbf{q}}{2}} - f_{m\mathbf{k}-\frac{\mathbf{q}}{2}}$, with $f_{n\mathbf{k}}$ being the Fermi occupation.
The energy difference $\hbar \omega_{\sigma}(\mathbf{q})=E_{n\mathbf{k}+\frac{\mathbf{q}}{2}} - E_{m\mathbf{k}-\frac{\mathbf{q}}{2}}$, with $E_{n\mathbf{k}}$ being the eigenvalue of the Bloch function $\psi_{n\mathbf{k}}$.
The light-matter interaction kernel $\mathcal{K}_{\sigma,ij}(\mathbf{q})=I_{\sigma,i}^{\dagger}(\mathbf{q})I_{\sigma,j}(\mathbf{q})$ with the interaction matrix $I_{\sigma,\bm{\epsilon}}(\mathbf{q})$ being expanded to the first order of $\mathbf{q}$ and omitting the spin part~\cite{Malashevich2010prb}
\begin{eqnarray}
    \label{eq:four}
    I_{\sigma,\bm{\epsilon}}(\mathbf{q}) &&= \left \langle \psi_{n\mathbf{k}+\frac{\mathbf{q}}{2}} \middle \vert \mathbf{J} \middle \vert \psi_{m\mathbf{k}-\frac{\mathbf{q}}{2}} \right \rangle \nonumber \\
    &&= \left \langle u_{n\mathbf{k}} \middle \vert \bm{\epsilon} \cdot \mathbf{v_k} \middle \vert u_{m\mathbf{k}} \right \rangle + \frac{\mathbf{q}}{2} \cdot \big \{ \nonumber \\
    && \left \langle \partial_{\mathbf{k}} u_{n\mathbf{k}} \middle \vert \bm{\epsilon} \cdot \mathbf{v_k} \middle \vert u_{m\mathbf{k}} \right \rangle - \left \langle u_{n\mathbf{k}} \middle \vert \bm{\epsilon} \cdot \mathbf{v_k} \middle \vert \partial_{\mathbf{k}} u_{m\mathbf{k}} \right \rangle \big \}
\end{eqnarray}
where $\bm{\epsilon}$ is the light polarization vector, $u_{n\mathbf{k}}(\mathbf{r})=\psi_{n\mathbf{k}}(\mathbf{r})e^{-i \mathbf{k} \cdot \mathbf{r}}$ is the cell-periodic part of the Bloch function. 
The paramagnetic current operator $\mathbf{J}=(e^{i\mathbf{q} \cdot \mathbf{r}} \mathbf{v} + \mathbf{v}e^{i\mathbf{q} \cdot \mathbf{r}} )/2$ with \textbf{v} being the velocity operator.
$\mathbf{v_k}=e^{-i\mathbf{k} \cdot \mathbf{r}}\mathbf{v}e^{i\mathbf{k} \cdot \mathbf{r}}$.
Using the relation $\mathbf{v_k}=dH_\mathbf{k}/\hbar d\mathbf{k}$, where the Hamiltonian $H_\mathbf{k}$ fulfills the Schr\"{o}dinger equation $H_\mathbf{k} u_{n\mathbf{k}} = E_{n\mathbf{k}} u_{n\mathbf{k}}$, and the manipulations in the Appendix, Eq.~(\ref{eq:four}) can be reformulated as
\begin{eqnarray}
    \label{eq:five}
    I_{\sigma,i}(q_l) =&& i \omega_{\sigma} P_{\sigma,i} \nonumber \\
    &&+ q_l \left( i \epsilon_{lij} M_{\sigma,j} - \omega_{\sigma} Q_{\sigma,il} + i \bar{v}_{\sigma,i} P_{\sigma,l}\right)
\end{eqnarray}
where $\epsilon_{lij}$ is the Levi-Civita symbol,  $\omega_\sigma = \omega_\sigma(\mathbf{q}=0)$, and $\bar{v}_{\sigma,i} = \partial_i(E_{n\mathbf{k}}+E_{m\mathbf{k}})/2\hbar$ with $\partial_i \equiv \partial_{k_i}$.
$P_{\sigma,i}$, $M_{\sigma,j}$, and $Q_{\sigma,il}$ are the components of the electric dipole, magnetic dipole, and electric quadrupole transition moments, respectively 
\begin{subequations}
    \label{eq:six}
    \begin{equation}
        \mathbf{P}_{\sigma} = i \left\langle u_{n\mathbf{k}} \middle\vert \partial_\mathbf{k} u_{m\mathbf{k}} \right\rangle
    \end{equation}
    \begin{equation}
        \label{eq:sixb}
        \mathbf{M}_{\sigma} = \frac{i}{4\hbar} \left\langle \partial_\mathbf{k} u_{n\mathbf{k}} \middle\vert \times \left(2H_\mathbf{k} - E_{n\mathbf{k}} - E_{m\mathbf{k}} \right) \middle\vert \partial_\mathbf{k} u_{m\mathbf{k}} \right\rangle
    \end{equation}
    \begin{equation}
        \label{eq:sixc}
        Q_{\sigma,ij} = \frac{i}{4} \left( \left\langle \partial_i u_{n\mathbf{k}} \middle\vert \partial_j u_{m\mathbf{k}} \right\rangle + \left\langle \partial_j u_{n\mathbf{k}} \middle\vert \partial_i u_{m\mathbf{k}} \right\rangle \right)
    \end{equation}
\end{subequations}
For finite systems, the last term of Eq.~(\ref{eq:five}) is zero and the multipole expansion formula is recovered~\cite{Malashevich2010prb}. 
For $n=m$, Eq.~(\ref{eq:sixb}) reduces to the orbital magnetic moment of Bloch electron~\cite{Xiao2010}. 
In general, Eqs.~(\ref{eq:sixb}) and (\ref{eq:sixc}) are gauge-dependent due to the $\partial_\mathbf{k} u_{n\mathbf{k}}$ term~\cite{Vanderbilt2018}. 
However, the gauge-dependence can be fixed by replacing the direct \textbf{k} derivative with covariant derivative, as detailed in a concurrent work done in coordination with the present one~\cite{Pozo-unpublished22}. 
Then with the sum-over-states formula, the individually gauge-covariant $M$ and $Q$ can be obtained 
\begin{subequations}
    \label{eq:seven}
    \begin{equation}
        \mathbf{M}_{\sigma}=\frac{i}{4\hbar} \sum_{p \neq n,m}  \left(2E_{p\mathbf{k}}-E_{n\mathbf{k}}-E_{m\mathbf{k}}\right) \left(\mathbf{P}_{\mathbf{k}np} \times  \mathbf{P}_{\mathbf{k}pm}\right)
    \end{equation}
    \begin{equation}
        Q_{\sigma,ij}=\frac{1}{4}\sum_{p \neq n,m} \left(P_{\mathbf{k}np,i}P_{\mathbf{k}pm,j}+P_{\mathbf{k}np,j}P_{\mathbf{k}pm,i}\right)
    \end{equation}
\end{subequations}
Eq.~(\ref{eq:seven}) is the sum-over-states formula for practical calculations of the magnetic dipole and electric quadrupole transition moments for periodic systems. Inserting Eq.~(\ref{eq:five}) into the interaction Kernel, we have 
\begin{eqnarray}
    \label{eq:eight}
    \mathcal{K}_{\sigma,ij}(q_l) = &&\omega_{\sigma}^2 A_{\sigma,ij} +q_l \big[ \omega_{\sigma} B_{\sigma,ijl} + \omega_{\sigma} C_{\sigma,ijl} \nonumber \\
    && + \omega_{\sigma} \left(\bar{v}_{\sigma,j}A_{\sigma,il}+\bar{v}_{\sigma,i}A_{\sigma,lj}\right) \big]
\end{eqnarray}
where
\begin{subequations}
      \label{eq:nine}
      \begin{equation}
          A_{\sigma,ij}=P_{\sigma,i}^\dagger P_{\sigma,j}
      \end{equation}
      \begin{equation}
          B_{\sigma,ijl} = \epsilon_{lji}P_{\sigma,i}^\dagger M_{\sigma,i} +\epsilon_{lij} P_{\sigma,j} M_{\sigma,j}^\dagger
      \end{equation}
      \begin{equation}
          C_{\sigma,ijl}=i \omega_\sigma (P_{\sigma,i}^\dagger Q_{\sigma,jl} - P_{\sigma,j} Q_{\sigma,il}^\dagger )
      \end{equation}
\end{subequations}
Inserting Eq.~(\ref{eq:eight}) into Eq.~(\ref{eq:three}) and differentiating the asymmetric part with respect to \textbf{q}, we finally obtain the optical activity tensor
\begin{equation}
    \label{eq:ten}
    \gamma_{ijl}(\omega) = \frac{e^2}{\Omega \varepsilon_0 \omega^2 \hbar}\sum_{\sigma} f_\sigma \left(g B_{\sigma,ijl}^{\prime \prime} +g C_{\sigma,ijl}^{\prime \prime} + D_{\sigma,ijl}\right)
\end{equation}
where $f_\sigma = f_\sigma(\mathbf{q}=0)$ and 
\begin{equation}
    D_{\sigma,ijl}=g\bar{v}_{\sigma,j}A_{\sigma,il}^{\prime \prime} + g\bar{v}_{\sigma,i}A_{\sigma,lj}^{\prime \prime} + g^2\bar{v}_{\sigma,l}A_{\sigma,ij}^{\prime \prime}
\end{equation}
The double primes denote the imaginary part. 
The first two terms \textit{B} and \textit{C} in the summand of Eq.~(\ref{eq:ten}) are the magnetic dipole and electric quadrupole contributions, respectively, while the last term \textit{D} is the band dispersion term which contains only the electric dipoles as in the multipole expansion.
$g=\omega_\sigma/(\omega-\omega_\sigma+i \eta)$ is spectral function.
Again, for finite systems, the band dispersion term is zero and Eq.~(\ref{eq:ten}) reduces to the formula for molecules. 

In general, the optical activity tensor has nine independent elements due to the antisymmetric relation $\gamma_{ijl} = -\gamma_{jil}$~\cite{Landau1984}. The nine elements can be rearranged into a second-rank axial tensor, the so-called gyration tensor $G_{ij}$
\begin{equation}
    G_{ij} = \frac{\omega}{2c}\epsilon_{iuv} \gamma_{uvj}
\end{equation}
Point group symmetry restricts the gyration tensor to particular form~\cite{Malgrange2014}. 
The optical activity tensor can be symmetrized accordingly based on the transformation rules.

From the definitions of the matrices in the summand of Eq.~(\ref{eq:ten}), it is clear that the ingredients needed to evaluate the optical activity tensor are the band energies $E_{n\mathbf{k}}$, band gradients or velocities $v_{n\mathbf{k}}$, and the electric dipole transition moments $P_\sigma$. 
The former can be easily obtained from normal density functional theory (DFT) calculations, while for the latter two there exist well-known approaches with different accuracy and efficiency~\cite{Gajdos2006,Sangalli2019}.
We implement the above formulation~\cite{q-optics} based on the outputs of DFT calculations via \textsc{vasp} code with the \texttt{LOPTICS} setting. 
The magnetic dipole and electric quadrupole transition moments are first calculated based on Eq.~(\ref{eq:seven}). 
Then, Eq.~(\ref{eq:ten}) is evaluated. 
Finally, the spectra of OR and CD can be obtained according to Eq.~(\ref{eq:two}).

\section{Application}

\subsection{OR of Te}
Elemental trigonal tellurium (Te) is the simplest chiral semiconductor with space group of $P3_121$ (right-handed) or $P3_221$ (left-handed).  
The unit cell contains only three tellurium atoms which covalently bonded to form a spiral chain along \textit{c} axis.  
The chains are arranged in hexagonal network with weak van der Waals interactions. 
The two enantiomorphic structures have same optical activity but opposite signs. 
We focus on the left-handed structure here. 
DFT calculations are performed with \textsc{vasp} code and PAW~\cite{Blochl1994} potentials.
Perdew-Burke-Ernzerhof~\cite{Perdew1996} (PBE) functional is used for the exchange-correlation interaction. 
An energy cutoff of 500 eV is used to expand the wave function. 
With spin-orbit coupling included, the calculated band gap is only 0.015 eV, to be compared to the experimental value of 0.323 eV~\cite{Anzin1977}. 
We employ a scissor shift~\cite{Fiorentini1995} to correct the band gap.

\begin{figure}[htbp]
\includegraphics[width=\linewidth]{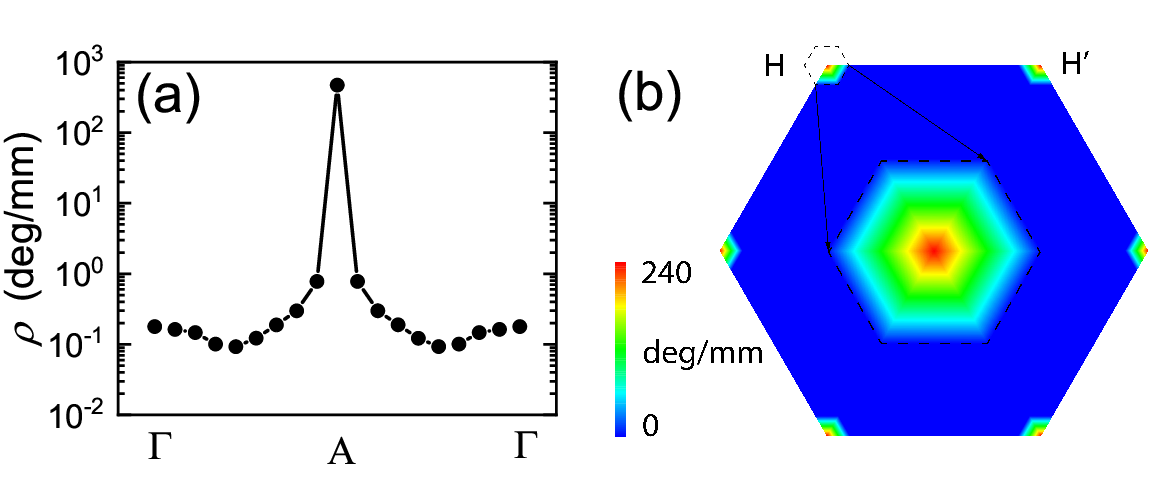}
\caption{\label{fig:1} $\mathbf{k}$-resolved OR of Te (a) along $\Gamma A$ and (b) in the 2D hexagonal BZ calculated at $\hbar \omega$ of 0.3 eV. The arrows and dashed lines indicate the enlargement of the BZ around $H$ point.}
\end{figure}

\begin{figure}[htbp]
\includegraphics[width=\linewidth]{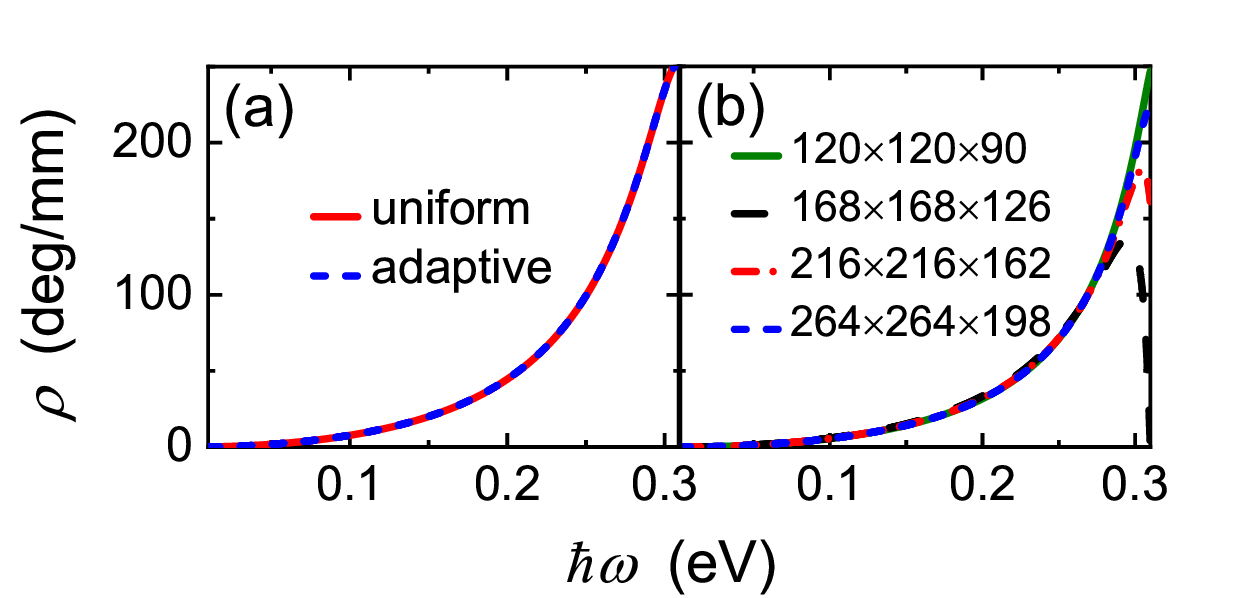}
\caption{\label{fig:2} (a) Comparison of the calculations on uniform and adaptive \textit{k} meshes. (b) Convergence of OR on the adaptive \textit{k} mesh.}
\end{figure}

To obtain converged OR of Te, extremely fine \textit{k} meshes are required for the Brillouin zone (BZ) integration. 
In previous implementation, this can be easily achieved by Wannier interpolation on a uniform $200 \times 200 \times 200$ \textit{k} mesh~\cite{Tsirkin2018}.
However, evaluating the required matrix elements on such fine \textit{k} mesh directly is very challenging. 
In this work, we use the adaptive \textit{k} mesh technique supplemented by symmetry reduction to reduce the computational challenge. 
We first analyze the \textbf{k}-resolved OR on a uniform $24 \times 24 \times 18$ \textit{k} mesh in the first BZ. 
Fig.~\ref{fig:1}(a) shows the plane-summed OR along the $\Gamma A$ direction. 
It is obvious that the plane containing $A (0, 0, 1/2)$ point dominates the contribution. 
If we resolve the \textbf{k} dependence in the $A$-contained 2D hexagonal BZ, we can observe that the OR is mainly contributed from the six corners, alternately denoted by $H(1/3, 1/3, 1/2)$ and $H^\prime (-1/3, -1/3, 1/2)$ which are related to each other with time-reversal symmetry, as shown in Fig.~\ref{fig:1}(b). 
This highly nonuniform $\mathbf{k}$ dependence of OR facilitates the application of the adaptive \textit{k} mesh technique. 
In our calculations, a denser \textit{k} mesh is used within a small box centered at $H$ (0.6\% of the entire BZ), while a coarser $24 \times 24 \times 18$ \textit{k} mesh is used elsewhere. 
Fig.~\ref{fig:2}(a) compares the OR calculated from the uniform and adaptive \textit{k} meshes. 
The results from the two \textit{k} meshes match exactly with each other. 
The number of irreducible \textit{k} points of the adaptive mesh is only 1174 compared to 46782 of the uniform mesh which indicates significant computation resource saving. 
We check the OR convergence on the adaptive \textit{k} mesh, as shown in Fig.~\ref{fig:2}(b). 
The lower energy part, i.e., less than 0.25 eV, can be easily converged with very few \textit{k} points. 
Challenges arise approaching the band gap where a resonant behavior is expected due to the denominators of the spectral functions $g$ and $h$.
Fine \textit{k} mesh as dense as $264 \times 264 \times 198$ is needed to obtain converged OR at around 0.31 eV.

To calculate the magnetic dipole and electric quadrupole transition moments, the sum-over-states formulation requires that the summations in Eq.~(\ref{eq:seven}) run over all the bands that determined by the Hamiltonian. 
In practice, however, the number of bands is truncated. 
As shown in Table~\ref{tab:1}, the OR is converged with 72 bands (9 valence bands).
\begin{table}[b]
    \caption{\label{tab:1}Convergence of the OR of Te as a function of number of bands (nbands). Data are in unit of deg/mm.}
    \begin{ruledtabular}
        \begin{tabular}{cccccc}
          \multirow{2}{*}{$\hbar \omega$ (eV)} & \multicolumn{3}{c}{nbands} & \multirow{2}{*}{Calc.\footnotemark[1]} & \multirow{2}{*}{Expt.\footnotemark[2]} \\
          & 18 & 36 & 72 & \\
          \hline
          0.117 & 8.3 & 7.9 & 7.8 & 4.9 & 9.0 $\pm$ 0.2 \\
          0.310 & 192.0 & 247.9 & 247.0 & & 145.2 $\pm$ 1.5
        \end{tabular}
    \end{ruledtabular}
    \footnotetext[1]{Wannier implementation~\cite{Tsirkin2018}.}
    \footnotetext[2]{Handness unknown~\cite{Fukuda1975}}
\end{table}
The calculated OR at 0.117 eV is 7.8 deg/mm which agrees well with experiment (9.0 deg/mm) but is larger than that of the Wannier implementation (4.9 deg/mm). 
The sign of the calculated OR is positive/negative for the left/right-handed Te, which is consistent with previous calculations~\cite{Tsirkin2018}.
However, there is difficulty in determining the handness of elemental crystals experimentally, thus conflicting claims about the absolute sign exist~\cite{Tsirkin2018}.
The OR calculated at 0.310 eV is 247.0 deg/mm which is 70\% larger than experiment. 
We will explain this later.

\begin{figure}[htbp]
\includegraphics[width=\linewidth]{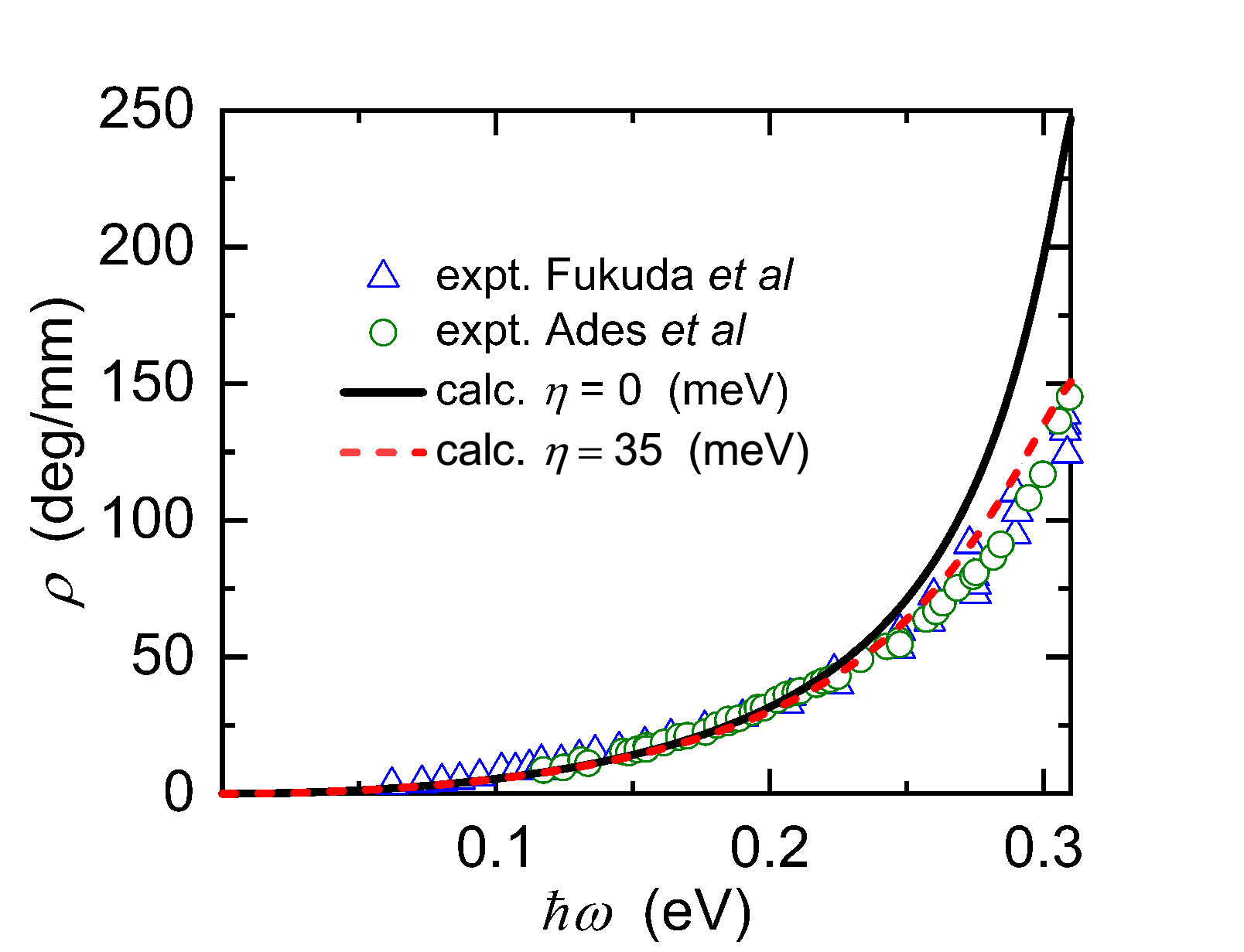}
\caption{\label{fig:3} Comparison of the calculated OR dispersion with experiments for Te.}
\end{figure}

The calculated OR dispersion or optical rotatory dispersion, compared with experiments, is shown in Fig.~\ref{fig:3}. 
The calculations agree with experiments quite well for the energy range less than 0.25 eV. 
Changing the broadening parameter $\eta$ hardly affect the dispersion at this lower energy range.
For higher energies that approaching the band gap, the OR dispersion is very sensitive to $\eta$. 
Since the experimental data are obtained at room temperature, thermal broadening effect comes to play an important role.  
A good agreement can be obtained with $\eta$ of 35 meV, as shown by the red dash line. 

\begin{figure}[htbp]
\includegraphics[width=\linewidth]{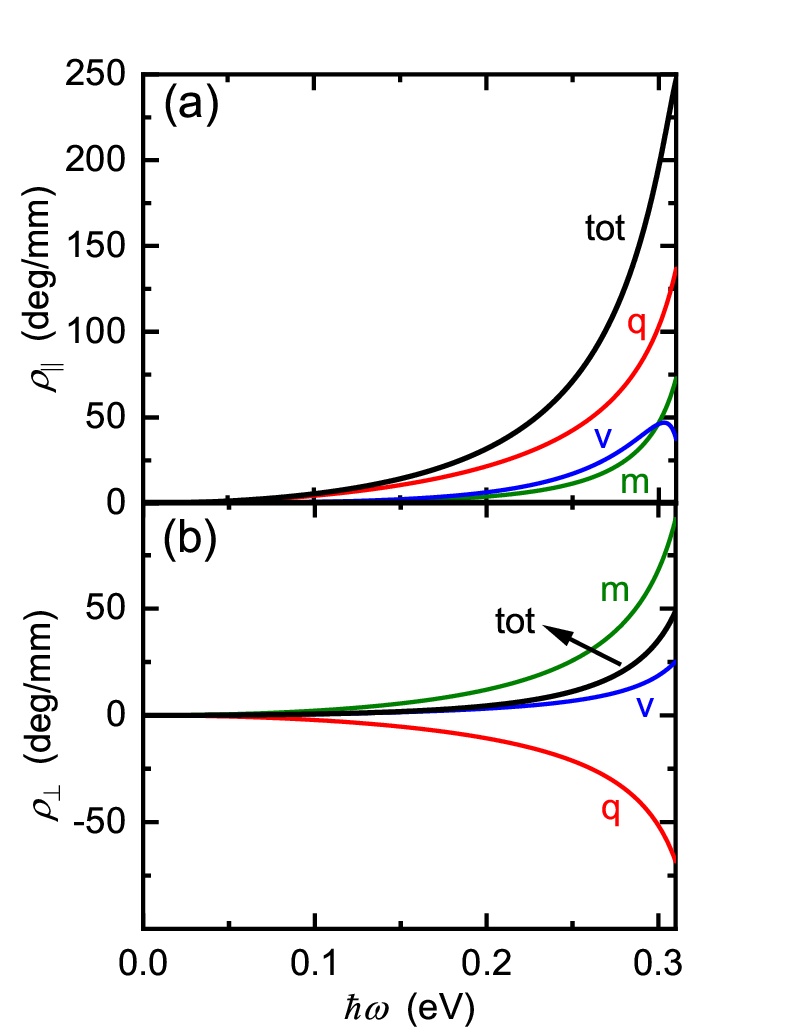}
\caption{\label{fig:4} OR of Te for the light propagating (a) along ($\rho_\parallel$) and (b) perpendicular to ($\rho_\perp$) the optic axis.}
\end{figure}

Since in experiments the OR in off optic axis is difficult to be measured~\cite{Malgrange2014,Kobayashi1978}, the available data for Te are all for light propagating along the optic axis. 
The symmetry operations of point group 32 reduce the OR tensor of Te to only two independent elements, i.e., $\rho_\parallel (\rho_{xyz})$ and $\rho_\perp (\rho_{yzx})$, for light propagating along and perpendicular to the optic axis. 
The calculated $\rho_\parallel$ is about five times larger than $\rho_\perp$, as shown in Fig.~\ref{fig:4}.
According to Eq.~(\ref{eq:ten}), the total OR can be decomposed into three different contributions.
We use \textit{m}, \textit{q}, and \textit{v} to denote the first, second, and third terms in the parenthesis of Eq.~\ref{eq:ten}, namely,  the magnetic dipole, electric quadrupole, and band dispersion terms, respectively. 
We calculate the different contributions for both $\rho_\parallel$ and $\rho_\perp$, as shown in Figs.~\ref{fig:4}(a) and (b), respectively.
For $\rho_\parallel$, the electric quadrupole term dominates the whole OR dispersion, contradicts to previous neglect of the electric quadrupole contribution to the optical activity. 
The band dispersion term shows a peak approaching the band gap.
This is attributed to the “camel back” band dispersion~\cite{Tsirkin2018} where the sign of the band gradient changes. 
For $\rho_\perp$, the magnetic dipole term and electric quadrupole term have opposite signs, leading to much reduced total OR. 
For both $\rho_\parallel$ and $\rho_\perp$, the band dispersion term makes a significant contribution, which is unique for calculating the optical activity of crystals.

\subsection{CD of chiral CNT}
Chiral CNTs with geometrical index $(n,m)$, where $n \neq m$ and $m \neq 0$, are a series of semiconductors that show obvious CD signal which can be used for enantiomer separation and electronic structure determination~\cite{Wei2016,Peng2007,Green2009}. 
Here, we focus on one of them, i.e., the $(6,4)$ CNT. 
The energy cutoff of 420 eV is used for the DFT calculations. 
We use $\eta$ of 0.1 eV to smear the CD spectra.
Unlike OR dispersion which is characterized at nonabsorbing frequencies, CD spectra are measured above the band gap. 
Hence, CD is more sensitive to the electronic band structures.
As semi-local density functionals like PBE would usually predict shrunk band width, we hereafter use HSE~\cite{Krukau2006,Heyd2003} functional for all the CD calculations.
It turns out that the CD spectra of $(6,4)$ CNT can be easily converged with a \textit{k} mesh of $1 \times 1 \times 32$ and only 360 total bands for evaluating Eq.~(\ref{eq:seven}).
Note that the number of valence bands is 304 for the unit cell. 

\begin{figure}[htbp]
\includegraphics[trim={0 8cm 0 0},clip,width=\linewidth]{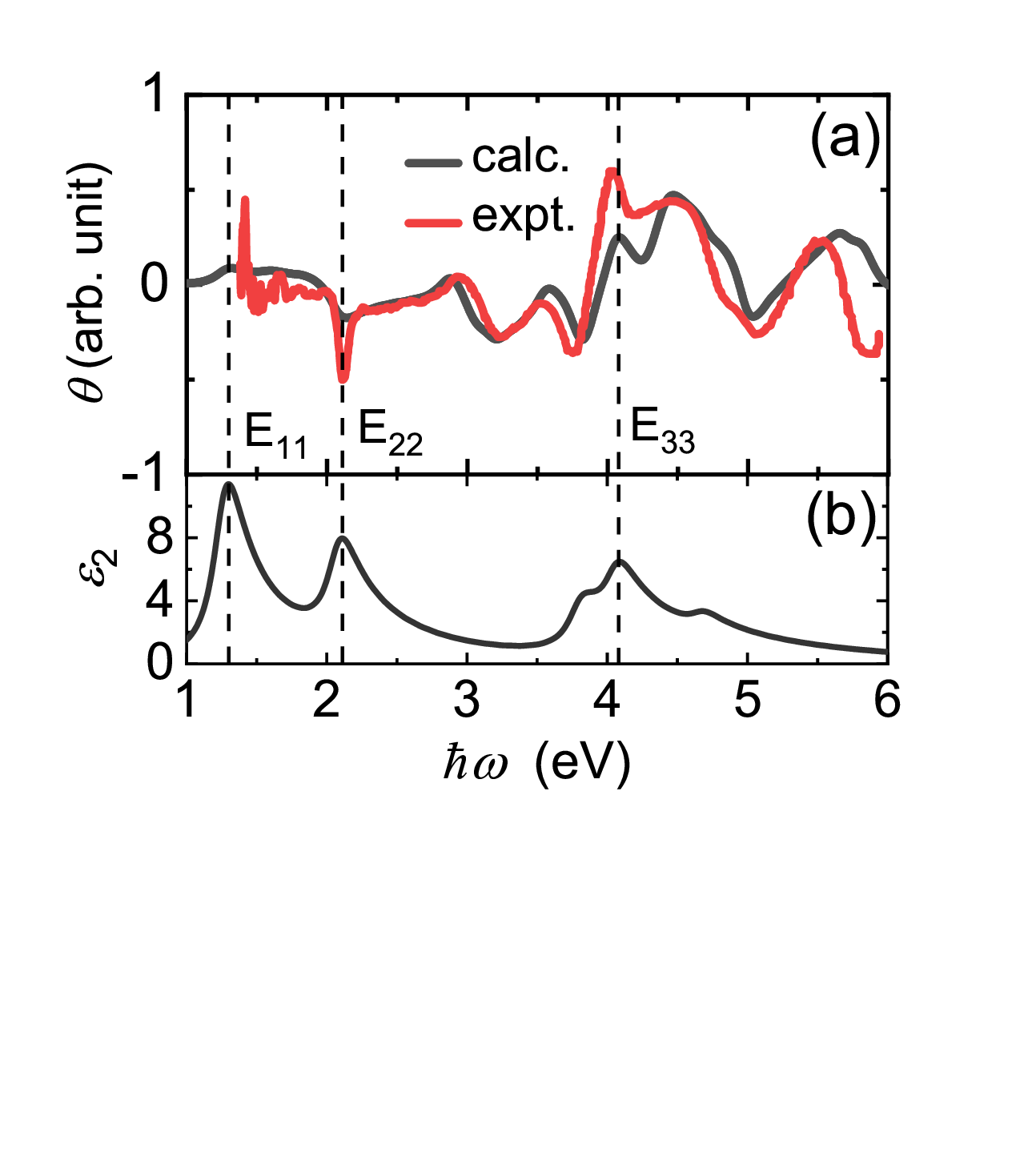}
\caption{\label{fig:5} (a) Comparison of the calculated CD spectra with experiment for (6,4) CNT. (b) The imaginary part of the dielectric function.}
\end{figure}

We compare the calculated CD spectra with experiment in Fig.~\ref{fig:5}(a). 
The calculated HSE band gap is 1.38 eV, which agrees with the interband transition $E_{11}$ of 1.42 eV as determined from the optical absorption spectra \cite{Wei2016}. 
Note, however, that this agreement is fortuitous due to the error cancellation, i.e., the underestimation of the band gap and the neglect of the excitonic effect~\cite{Wagner2019}. 
To make better comparison, we align the $E_{22}$ transition peaks in the CD spectra, as the lower energy part around the $E_{11}$ transition shows much more noise~\cite{Wei2016}. 
The calculated CD spectra reproduce all the key features of the experimental data. 
The main discrepancy comes from the underestimated intensities of the $E_{11}$, $E_{22}$, and $E_{33}$ peaks, i.e., the main absorption peaks as indicated in Fig.~\ref{fig:5}(b), which is attributed to the excitonic effect that neglected in the current formulation.

\begin{figure}[htbp]
\includegraphics[width=\linewidth]{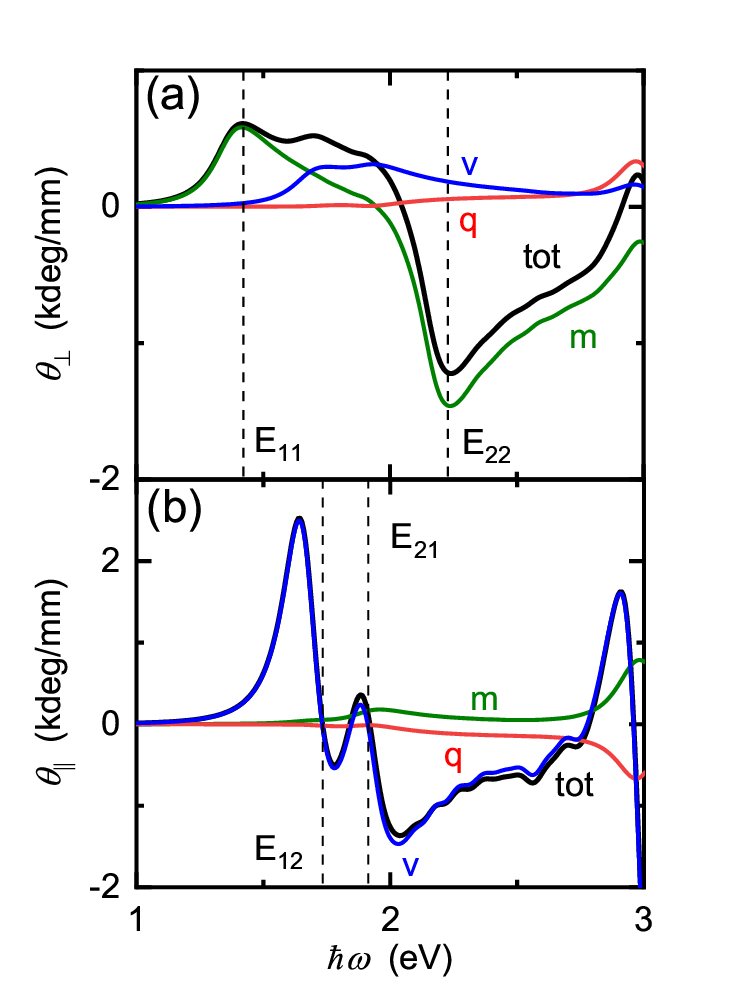}
\caption{\label{fig:6} CD of $(6,4)$ CNT for light propagating (a) perpendicular to ($\theta_\perp$) and (b) along ($\theta_\parallel$) the tube axis.}
\end{figure}

The gyration tensor of $(6,4)$ CNT has two independent elements, i.e., $G_{11} = G_{22}$ and $G_{33}$~\cite{Damn1999}. 
We denote the anisotropic CD by $\theta_\parallel$ and $\theta_\perp$ for the light propagating along and perpendicular to the tube axis, respectively, as shown in Fig.~\ref{fig:6}. 
We find that $\theta_\parallel$ is larger than $\theta_\perp$ for the whole spectrum. 
In experiment, the CD was measured for solutions of CNTs with random orientation, for which $\theta_\perp$ dominates~\cite{Wei2016}.
Therefore, measuring $\theta_\parallel$ is still challenging.

We further decompose the CD spectra into different contributions in Figs.~\ref{fig:6}(a) and (b). 
For $\theta_\perp$, the $E_{11}$ and $E_{22}$ peaks are almost exclusively contributed from the magnetic dipole term. 
The band dispersion term dominates the $E_{12}$ and $E_{21}$ peaks, which is not present in previous calculations~\cite{Sanchez2010}, while the electric quadrupole term arises only beyond the $E_{22}$ peak.
As for $\theta_\parallel$, the band dispersion term dominates the whole spectrum. The magnetic dipole term and the electric quadrupole term have opposite signs. Their contributions are small for the lower energy range ( $<$ 3 eV) and almost cancel with each other for higher energies (not shown). 
There is no CD signal around $E_{11}$ for $\theta_\parallel$ due to the optical selection rule that only $E_{nm}$ transitions are allowed for light propagating along the tube axis.
Although the CD spectra at $E_{12}$ and $E_{21}$ are mainly derived from the band dispersion term for both $\theta_\parallel$ and $\theta_\perp$, the band shapes are quite different, i.e., monosignate for $\theta_\perp$ versus bisignate for $\theta_\parallel$.
While monosignate line shape for a specific optical transition is quite common for CD spectra, bisignate line shapes can also be observed for multiple chromophores with chiral alignment~\cite{Berova2007}.
Here, the bisignate band shape of $\theta_\parallel$ is, however, attributed to a different origin.
The band dispersion term of Eq.~(\ref{eq:ten}) $D_{\sigma,ijl}$ has three components with band velocities along $i,j$ and $l$, respectively.
Due to the quasi-one dimensional nature of $(6,4)$ CNT, only the band velocity along the tube axis is nonzero, i.e., $\bar{v}_{\sigma,x}=\bar{v}_{\sigma,y}=0$. 
For $\theta_\perp$, the corresponding tensor components are $yzx$ and $zxy$. 
Either the first or second component of $D_{\sigma,ijl}$ with spectral function of $g$ is nonzero.
The tensor component of $\theta_\parallel$ is $xyz$, hence only the last component with spectral function of $h$ survives.
Since $g^{\prime \prime} \sim \delta$ and $h^{\prime \prime} \sim \delta^\prime$ with $\delta$ and $\delta^\prime$ being the Dirac delta function and its derivative are monosignate and bisignate, respectively, the corresponding band shapes of the CD spectra result.  

\subsection{Optical activity of achiral crystals}
Although all chiral crystals are expected to be optically active, being chiral is not a necessary condition for optical activity. 
Many achiral crystals have shown optical rotatory power~\cite{Claborn2008}. 
Experimentally, OR is determined by the symmetrical part of the gyration tensor~\cite{Malgrange2014}, which can also be nonzero for crystals with achiral $m$, $mm2$, $\bar{4}$, and $\bar{4}2m$ point groups. 
One typical and well-known example is $\textrm{AgGaS}_2$~\cite{Hobden1967,Hobden1968}, which has a point group of $\bar{4}2m$. 
For the point groups of $3m$, $4mm$, and $6mm$, there is only one independent gyration tensor element, i.e., $G_{12} = -G_{21}$. 
In general, crystals with gyrotropic point groups can be called optically active~\cite{Malgrange2014}. 
The difference between the optical activity of materials with chiral/achiral gyrotropic groups is that the trace of the gyration tensor is nonzero/zero.
For chiral groups, optical activity can be observed for light propagating along any directions and survive after isotropic averaging, while for achiral groups, light incident directions are restricted.
As an example, we calculate the CD of wurtzite GaN which has a point group of $6mm$. 
DFT calculations are performed with energy cutoff of 400 eV. 
The calculated PBE band gap is 1.715 eV.
Adaptive \textit{k} mesh with 0.7\% of the BZ centered at the $\Gamma$ point being refined by a $120 \times 120 \times 75$ mesh and $24 \times 24 \times 15$ elsewhere is used.
The number of total bands of 36 converges the summation in the sum-over-states formulation.  

\begin{figure}[htbp]
\includegraphics[width=\linewidth]{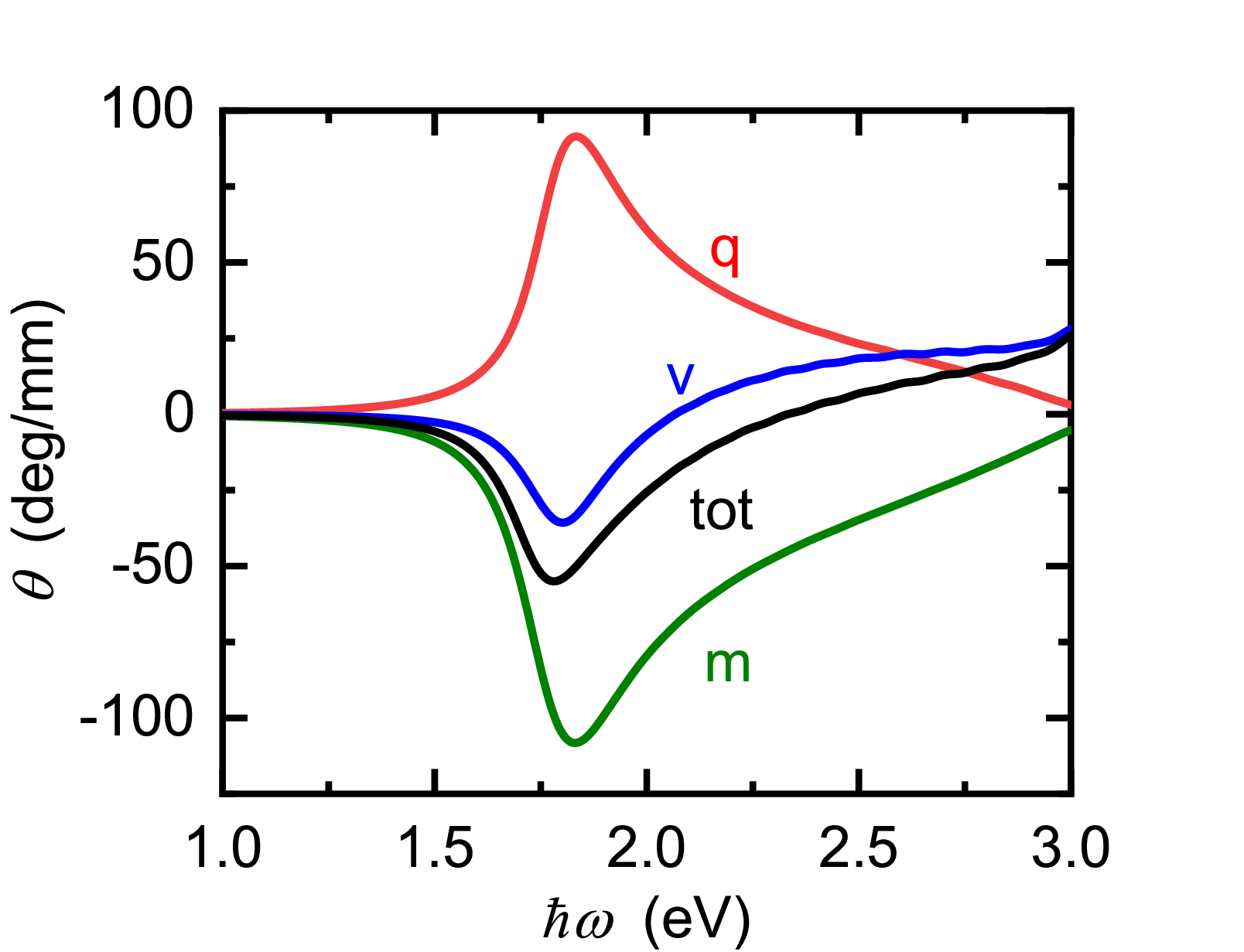}
\caption{\label{fig:7} Optical activity of wurtzite GaN.}
\end{figure}

The nonzero component $\theta_{yzy}$ with different contributions are shown in Fig.~\ref{fig:7}.  
Again, the magnetic dipole and electric quadrupole terms have opposite signs. 
The mutual cancellation significantly reduces the total CD, of which more than 65\% is contributed from the band dispersion term.
Our calculations confirm that achiral crystals with gyrotropic point groups can also be optically active. 
Indeed, the component $\theta_{yzy}$ should be related to the longitudinal excitons as the the electric field and wave vector are in the same direction.
Although how to measure the CD of such achiral crystals is beyond the scope of the present work, great diversity of optically active materials with achiral point groups may lead to future researches and findings thereof. 

\section{Summary}
From the expansion of the light-matter interaction Hamiltonian, we formulate the optical activity with different contributions: the magnetic dipole term, the electric quadrupole term, and the band dispersion term, with the last one being unique for crystals. 
The magnetic dipole and electric quadrupole transition moments for periodic systems can be calculated with a sum-over-states formulation.
We apply our formulation to calculate the OR of Te and CD of $(6,4)$ CNT, respectively, demonstrating the computationally challenging and friendly cases. 
The former requires very dense \textit{k} points for the BZ integration, which is remedied with the adaptive \textit{k} mesh technique, and many conduction bands for the sum-over-states formulation, while the latter needs very few \textit{k} points and bands. 
From the decomposed optical activity, the signs of the magnetic dipole term and the electric quadrupole term can be same or opposite.
For both of the systems, the band dispersion term plays an important role. 
In particular, for $(6,4)$ CNT, the calculated CD for light propagating along the tube axis is contributed almost exclusively from the band dispersion term and is much larger than that of perpendicular incidence.
Good agreements with experiments are obtained for the calculated OR dispersion and CD spectra.
Finally, we calculate the CD of wurtzite GaN as a showcase for achiral crystals, confirming the optical activity in achiral gyrotropic point groups.

\begin{acknowledgments}
This work was supported by the Center for Hybrid Organic Inorganic Semiconductors for Energy (CHOISE), an Energy Frontier Research Center funded by the Office of Basic Energy Sciences, Office of Science within the US Department of Energy. 
The research was performed using computational resources sponsored by the Department of Energy's Office of Energy Efficiency and Renewable Energy and located at the National Renewable Energy Laboratory.
Y.Y. acknowledges the support of National Science Foundation under contract no. DMR-1807818.
We thank Stepan Tsirkin, Ivo Souza for useful comments on the manuscript and fruitful discussions. 
\end{acknowledgments}

\appendix*
\section{}
We show here how to obtain Eq.~(\ref{eq:five}) from Eq.~(\ref{eq:four}) in the main text. 
Assume that the light wave vector and polarization directions are \textit{l} and \textit{i}, respectively, we have for the first term of Eq.~(\ref{eq:four}) 
\begin{eqnarray}
    \left\langle u_n \middle \vert v_i \middle \vert u_m \right\rangle &&= i \frac{E_n-E_m}{\hbar} \left\langle u_n \middle \vert r_i \middle \vert u_m \right\rangle \nonumber \\
    &&= i \omega_{nm} \left\langle u_n \middle \vert \partial_i u_m \right\rangle = i \omega_{nm} P_{nm,i}
\end{eqnarray}
where $\hbar \omega_{nm} = E_n -E_m$. We used $\mathbf{v}=i /\hbar [\mathbf{H}, \mathbf{r}]$ and $\mathbf{r} = i \partial_\mathbf{k}$.
And the second term of Eq.~(\ref{eq:four})

\begin{eqnarray}
    \label{eq:A1}
    &&\left\langle \partial_l u_n \middle \vert \partial_i H \middle \vert u_m \right\rangle \nonumber \\
    =&& \partial_i \left\langle \partial_l u_n \middle \vert H \middle \vert u_m \right\rangle - \left\langle \partial_i \partial_l u_n \middle \vert H \middle \vert u_m \right\rangle - \left\langle \partial_l u_n \middle \vert H \middle \vert \partial_i u_m \right\rangle \nonumber \\
    =&& \partial_i \left\langle \partial_l u_n \middle \vert E_m \middle \vert u_m \right\rangle - \left\langle \partial_i \partial_l u_n \middle \vert E_m \middle \vert u_m \right\rangle - \left\langle \partial_l u_n \middle \vert H \middle \vert \partial_i u_m \right\rangle \nonumber \\
    =&& \left\langle \partial_i \partial_l u_n \middle \vert E_m \middle \vert u_m \right\rangle + \left\langle \partial_l u_n \middle \vert \partial_i E_m \middle \vert u_m \right\rangle + \left\langle \partial_l u_n \middle \vert E_m \middle \vert \partial_i u_m \right\rangle \nonumber \\
    &&- \left\langle \partial_i \partial_l u_n \middle \vert E_m \middle \vert u_m \right\rangle - \left\langle \partial_l u_n \middle \vert H \middle \vert \partial_i u_m \right\rangle \nonumber \\
    =&& i \partial_i E_m P_{nm,l} + \left\langle \partial_l u_n \middle \vert E_m - H \middle \vert \partial_i u_m \right\rangle
\end{eqnarray}

Similarly the last term

\begin{eqnarray}
    \label{eq:A2}
    &&\left\langle u_n \middle \vert \partial_i H \middle \vert \partial_l u_m \right\rangle \nonumber \\
    =&& \left\langle \partial_i u_n \middle \vert E_n - H \middle \vert \partial_l u_m \right\rangle - i \partial_i E_n P_{nm,l}
\end{eqnarray}

Take the difference of Eqs.~(\ref{eq:A1}) and (\ref{eq:A2}) 

\begin{eqnarray}
    \label{eq:A3}
    &&\left\langle \partial_l u_n \middle \vert E_m - H \middle \vert \partial_i u_m \right\rangle - \left\langle \partial_i u_n \middle \vert E_n - H \middle \vert \partial_l u_m \right\rangle \nonumber \\
    && + i \partial_i E_m P_{nm,l} + i \partial_i E_n P_{nm,l} \nonumber \\
    =&& -\left( \left\langle \partial_l u_n \middle \vert H \middle \vert \partial_i u_m \right\rangle - \left\langle \partial_i u_n \middle \vert H \middle \vert \partial_l u_m \right\rangle \right) \nonumber \\
    && + \frac{1}{2} \left( \left\langle \partial_l u_n \middle \vert E_m \middle \vert \partial_i u_m \right\rangle - \left\langle \partial_i u_n \middle \vert E_m \middle \vert \partial_l u_m \right\rangle \right) \nonumber \\
    && + \frac{1}{2} \left( \left\langle \partial_l u_n \middle \vert E_m \middle \vert \partial_i u_m \right\rangle + \left\langle \partial_i u_n \middle \vert E_m \middle \vert \partial_l u_m \right\rangle \right) \nonumber \\
    && + \frac{1}{2} \left( \left\langle \partial_l u_n \middle \vert E_n \middle \vert \partial_i u_m \right\rangle - \left\langle \partial_i u_n \middle \vert E_n \middle \vert \partial_l u_m \right\rangle \right) \nonumber \\
    && - \frac{1}{2} \left( \left\langle \partial_l u_n \middle \vert E_n \middle \vert \partial_i u_m \right\rangle + \left\langle \partial_i u_n \middle \vert E_n \middle \vert \partial_l u_m \right\rangle \right) \nonumber \\
    && + i \partial_i (E_n + E_m) P_{nm,l} \nonumber \\
    =&& - \frac{1}{2} \big[ \left\langle \partial_l u_n \middle \vert 2H-E_n - E_m \middle \vert \partial_i u_m \right\rangle \nonumber \\
        && \quad \quad \quad \quad - \left\langle \partial_i u_n \middle \vert 2H-E_n - E_m \middle \vert \partial_l u_m \right\rangle \big ] \nonumber \\
    &&-\frac{E_n - E_m}{2} \left[ \left\langle \partial_l u_n \middle \vert  \partial_i u_m \right\rangle + \left\langle \partial_i u_n \middle \vert  \partial_l u_m \right\rangle \right] \nonumber \\
    && + i \partial_i (E_n + E_m) P_{nm,l}
\end{eqnarray}
The first and second terms in Eq.~(\ref{eq:A3}) are the magnetic dipole and electric quadrupole matrix elements, respectively, as shown in Eqs.~(\ref{eq:sixb}) and (\ref{eq:sixc}).

\bibliography{refs}
\bibliographystyle{apsrev}

\end{document}